# Task-driven Self-supervised Bi-channel Networks for Diagnosis of Breast Cancers with Mammography

Ronglin Gong
School of Communication and Information Engineering
Shanghai University
Shanghai, China
gongronglin@shu.edu.cn

Jun Wang
School of Communication and Information Engineering
Shanghai University
Shanghai, China
wangjun_shu@shu.edu.cn

Jun Shi*
School of Communication and Information Engineering
Shanghai University
Shanghai, China
junshi@shu.edu.cn

*Abstract*—Deep learning can promote the mammography-based computer-aided diagnosis (CAD) for breast cancers, but it generally suffers from the small sample size problem. Self-supervised learning (SSL) has shown its effectiveness in medical image analysis with limited training samples. However, the network model sometimes cannot be well pre-trained in the conventional SSL framework due to the limitation of the pretext task and fine-tuning mechanism. In this work, a Task-driven Self-supervised Bi-channel Networks (TSBN) framework is proposed to improve the performance of classification model the mammography-based CAD. In particular, a new gray-scale image mapping (GSIM) is designed as the pretext task, which embeds the class label information of mammograms into the image restoration task to improve discriminative feature representation. The proposed TSBN then innovatively integrates different network architecture, including the image restoration network and the classification network, into a unified SSL framework. It jointly trains the bi-channel network models and collaboratively transfers the knowledge from the pretext task network to the downstream task network with improved diagnostic accuracy. The proposed TSBN is evaluated on a public INbreast mammogram dataset. The experimental results indicate that it outperforms the conventional SSL and multi-task learning algorithms for diagnosis of breast cancers with limited samples.

## I. Introduction

X-ray mammography is a routine imaging tool for diagnosis of breast cancers [1]. The mammography-based computer-aided diagnosis (CAD) has shown its effectiveness with good repeatability and consistency [2].

As a classical deep learning (DL) method, convolution neural networks (CNN) have shown their effectiveness in the field of mammography-based CAD [3][4]. However, it is often expensive and time-consuming to collect and annotate a large number of mammograms. Consequently, the limited training samples cannot well train an accurate and stable DL model [5]. Therefore, it is still a challenging task to improve the diagnosis accuracy of a mammography-based CAD with limited data.

In recent years, self-supervised learning (SSL) has attracted considerable attention as an effective way for few-shot learning [6]. It aims to automatically leverage the inherent information of data itself as supervision for the specific pretext task with pseudo labels [7]. In the SSL framework, a pre-trained network model is first acquired by this pretext task, and the learned model parameters are then transferred to the downstream task. The downstream model is fine-tuned with limited training samples for more stable and superior performance [6]. Since the pretext task is performed on the training samples generated from the downstream task, it is unnecessary for the SSL model to be pre-trained on a large amount of additional data for adaptation. Therefore, SSL is suitable for different tasks of medical image analysis with small sample size.

Several pretext tasks have been proposed in the early SSL works, such as jigsaw puzzles and image inpainting [8][9]. They are designed by utilizing the intrinsic properties of image samples, and the network model thus learns corresponding feature representation during this pre-training stage, such as spatial structures or color information, which benefits the downstream target task. That is to say, the design of the pretext task greatly affects the performance of the downstream task model.

Some pioneering SSL works on medical image analysis have also indicated the effectiveness of reasonably designed tasks [10][11]. Among them, image restoration is commonly utilized as the pretext task for medical images, because it fully uses the medical data, and the network model can learn rich context knowledge [11][12]. For example, Chen et al. developed the image context restoration task by restoring the exchanged patches with random positions in the original medical image, and the pre-trained model was then applied to the classification, localization, and segment tasks for medical images [11]; Zhou et al. proposed the Models Genesis as a pretext task that applied one or more transformations to 3D medical images, and the downstream model trains a segmentation model with superior generalization performance [12]. These works suggest the promising prospect of restoration-based tasks for medical imaging data.

It is worth noting that in the conventional SSL framework, the pseudo labels are self-generated without manual annotation for the pretext task, while the original labeled training samples are utilized for downstream supervised learning. Thus, the original label information is generally ignored in the pretext task. However, it has been proven that the SSL tasks using the label information gain more competitive performance [13]. In fact, since CADs, including mammography-based CAD, generally performs binary classification or a classification task with limited categories, it is feasible to embed the class information of a CAD task into a pretext image restoration task, which has the potential to

---

promote the pretext task network to learn more discriminative feature representation.

On the other hand, since the transfer method of SSL is based on fine-tuning, the networks for both pretext and downstream tasks have the same backbone. However, the pre-trained model sometimes cannot fully learn the feature representation from the pretext task in this fine-tuning based framework, because different networks perform differently across the specific pretext tasks [14]. For example, the ResNet model can perform the classification tasks well [15], but it generally cannot efficiently learn the hierarchical features about training samples from any image restoration task, since the ResNet architecture is not specifically designed for this kind of task. Consequently, it is necessary to further improve the flexibility of the current SSL framework. Specifically, a feasible method is to develop two different network models for the pretext and downstream tasks, respectively.

It is known that deep multi-task learning utilizes common representation across multiple related tasks to improve the performance of DL model [16]. It generally conducts hard or soft parameter sharing of the partial network model, which transfers knowledge between different tasks through simultaneously learning [17]. Inspired by this framework, it is feasible that two different networks can also be jointly trained by the pretext and downstream tasks, respectively.

To this end, we propose a novel Task-driven Self-supervised Bi-channel Networks (TSBN) framework for diagnosis of breast cancers with mammography. A new image restoration pretext task, named gray-scale image mapping (GSIM), is developed by embedding label information into the image restoration task. Moreover, inspired by the conventional SSL and multi-task learning, our framework jointly trains two different networks for the pretext image restoration task and downstream classification task, which promotes the classification model with collaboratively transferred knowledge. The experimental results on the INbreast dataset indicate the effectiveness of the proposed TSBN for the mammography-based CAD model with limited training samples.

The main contributions of this work are two-fold:

- A novel GSIM that uses original label information is proposed as a pretext image restoration task. In particular, the labels of binary classification is embedded into this task, and therefore, the learned feature representation contains discriminative information on benign and malignant mammograms, which improves the diagnostic accuracy of the downstream task network for breast cancers.
- A new TSBN framework is developed, which is totally different from the existing SSL framework based on fine-tuning. It integrates two different networks into a unified SSL framework for both the pretext and downstream tasks, and collaboratively transfers the learned knowledge from the pretext task to the downstream task. Since the pretext task is specifically developed for the proposed GSIM, it can learn more effective feature representation, which then improves the performance of the downstream classification network for mammography-based CAD.

II. METHOD

Figure 1 shows the flowchart of the proposed TSBN framework that contains three modules: pretext task channel module, downstream task channel module, and collaborative transfer module.

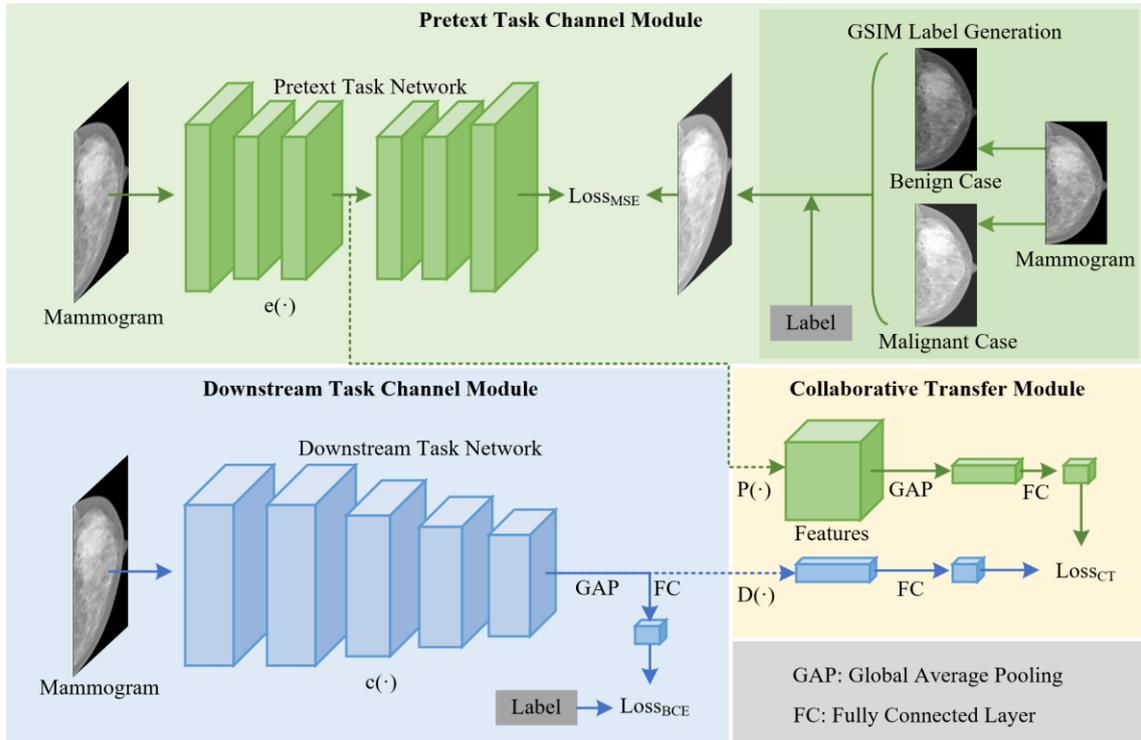

Fig. 1. Flowchart of task-driven self-supervised bi-channel networks, where different network architectures are adopted for the pretext and downstream tasks, respectively. $e(\cdot)$ is the encoder of the pretext task network followed by transfer branch $P(\cdot)$, which transfers the knowledge learned on gray-scale image mapping (GSIM) task, while $c(\cdot)$ is the encoder of the downstream task network followed by transfer branch $D(\cdot)$, which receives the knowledge from the pretext task channel module.

A new GSIM task is designed in the pretext task channel module, which guides CNN to learning more discriminative feature representation from the image restoration task. The bi-channel networks are jointly trained, and the learned feature representations in both networks are collaboratively transferred to each other by the collaborative transfer module. Specifically, TSBN is alternately trained with the following two steps:

- In the pretext task step, the mammograms are sent to the pretext task network, to generate the target images with category-related information.
- In the downstream task step, the mammograms are fed into the pretext and downstream task network, respectively. The downstream task network produces the classification results, and both networks generate the features to be input to the collaborative transfer module, which is utilized for knowledge transfer by minimizing the discrepancy with the pretext channel features.

Thus, the performance of the classification network is improved in this new SSL framework, which can effectively promote diagnostic accuracy for breast cancers. Then, the trained downstream network is utilized as the mammography-based CAD model to predict the new mammograms independently.

*A. Pretext Task Channel Module*

**Pretext Task Based on Gray-Scale Image Mapping.** The mammography-based CAD for breast cancers conducts a binary classification task, which makes it feasible to embed the label information of class into the pseudo label pixels in an image restoration task. We design a novel image-to-image self-supervised transform task, namely GSIM, as the pretext task in this module. This new task can make the specifically designed pretext task network to learn more effective basic representation and additional class information about mammograms.

As shown in Fig. 1, the GSIM task serves the original mammograms as the input of the pretext task network, and utilized the transformed mammograms as the pseudo labels, in which the gray-scale value of each pixel is mapped according to the label information of binary classification task. Formally, the transformation function of the GSIM label generation is defined as:

$$F(t) = t - (-1)^a d/2 \,, \ t \in [0,1] \quad (1)$$

where $t$ is a pixel value in the mammogram after normalization, $a \in \{0,1\}$ denotes the category of the sample, with 0 as the benign case and 1 as the malignant case. $d$ represents the distance between the two candidate target tensors corresponding to the two classes, and we select one tensor as the target of the pretext task network according to the label.

Mean squared error is adopted as $Loss_{MSE}$ to train the network in the pretext task channel module. The network model learns to map the original mammograms into the target images with label information. This training process can implicitly promote the pretext task network to learn the category-related discriminative features from this image-to-image task.

It is worth noting that in the conventional SSL framework, the backbone of the downstream task network is pre-trained by the automatically synthesized samples. The data synthesis methods generally cannot build a close correlation between the pretext and downstream tasks, which sometimes affects the transfer performance. While in our GSIM task, the category-related information is embedded into the pseudo label, thus this new pretext task has a prominent correlation with the downstream task, which also provides promising compatibility during our bi-channel networks training.

**Pretext Task Network.** Since the architecture design of the classification networks leaves out fine-grained features and focuses on learning discriminative semantic features in deep layers [18], it is not suitable for the image restoration tasks, and cannot fully learn the feature representation from our GSIM task. Therefore, it is necessary to develop or adopt the specifically designed CNN for image restoration tasks. Since the purpose of this work mainly evaluates the effectiveness of the proposed TSBN framework, we adopt two different CNN architectures, namely U-Net [19] and residual dense network (RDN) [20], to perform the proposed GSIM tasks, respectively.

U-Net is a well-known encoder-decoder architecture that is originally designed for medical image segmentation [19]. This network architecture and its variants can also effectively perform image restoration tasks that generally restores or reconstructs the information of high-quality images [21]. As shown in Fig. 2, the output features of the U-Net encoder are used for transfer in this work. Since U-Net is widely used, please refer to [19] for more related information.

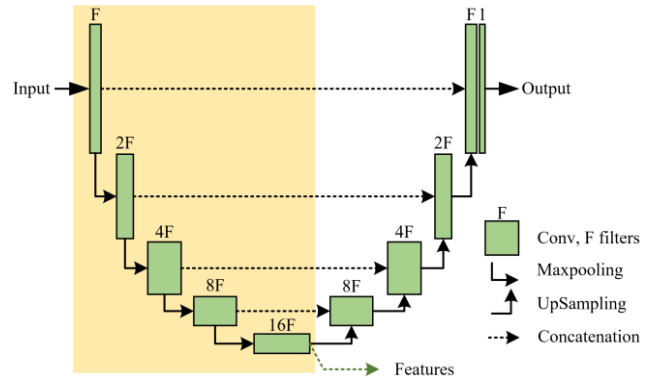

Fig. 2. Architecture of U-Net, where the partial network in the block of a light yellow background is regarded as the U-Net encoder $e(\cdot)$.

RDN has been successfully applied to image denoising and super-resolution tasks [20]. It also has the feasibility to well conduct the GSIM task. As shown in Fig. 3, RDN introduces the proposed cascaded residual dense blocks (RDB) in the network architecture to learn and fuse the hierarchical features, which takes both advantages of residual blocks and densely connected convolutional layers. In this work, the output features of the network encoder, which is shown in the block with a light yellow background in Fig. 3, will be transferred to the downstream task network. Please refer to [20] for more related information about RDN.

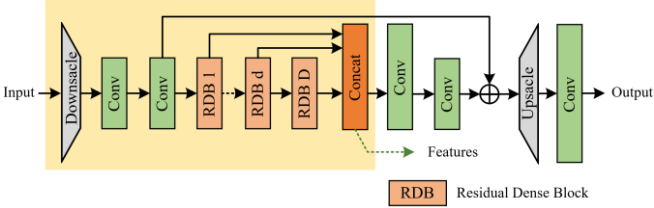

Fig. 3. Architecture of residual dense network (RDN), where the partial network in the block of a light yellow background is regarded as the RDN encoder $e(\cdot)$.

## B. Downstream Task Channel Module

The target task of this module is to perform a mammography-based CAD. Specifically, given a mammogram, the downstream task network tries to discriminate the image into the benign or malignant case. Since ResNet has shown its great capability for mammography-based CAD of breast cancers in previous work [22][24], we utilize it as the backbone network in this module. The output features, which are extracted from the backbone network followed by global average pooling are sent into the collaborative transfer module for knowledge transfer. Please refer to [15] for more related information about the design of ResNet architecture.

It is also worth noting that the downstream task network should be specifically designed for different downstream tasks.

## C. Collaborative Transfer Module

In TSBN, two different networks are respectively used in the pretext channel and downstream task channel modules. Therefore, the conventional transfer methods, such as fine-tuning and multi-task learning that uses a shared backbone, are not applicable in our work. To this end, a collaborative transfer loss is adopted to conduct knowledge transfer between the bi-channel networks in this module.

As shown in Fig. 1, $e(\cdot)$ is the encoder of the pretext task network followed by the branch $P(\cdot)$, which transfers the learned features to downstream channel module, while $c(\cdot)$ is the encoder of the downstream network followed by the branch $D(\cdot)$, which receives the features from the pretext task channel module. The collaborative loss function is formalized by:

$$Loss_{CT} = \sum(P(e(x_i)) - D(c(x_i)))^2 \quad (2)$$

where $x_i$ denotes the $i$-th mammogram sample.

In addition, we utilize the binary cross-entropy loss with category weight as follows:

$$Loss_{BCE} = -\sum w y_i log \hat{y}_i + (1 - y_i) log(1 - \hat{y}_i) \quad (3)$$

where $y_i$ denotes the $i$-th label, $\hat{y}_i$ is the prediction of the $i$-th sample. $w$ is the weight parameter, which is set to 15 in this work for class imbalance.

Considering that the knowledge is collaboratively transferred to improve the classification performance of the model, we define the total loss function of the downstream task step as:

$$Loss_D = \alpha Loss_{CT} + Loss_{BCE} \quad (4)$$

where $\alpha$ determines the coefficients of $Loss_{CT}$ and $Loss_{BCE}$.

In fact, both pretext and downstream tasks adopt the same training samples to train two different CNNs. In order to efficiently learn intrinsic representation from the limited samples for both networks, the bi-channel networks are jointly trained with this collaborative transfer module instead of the conventional fine-tuning strategy. Here, the parameters of the pretext task network and downstream task network are alternately trained by $Loss_{MSE}$ and $Loss_D$, respectively. Consequently, the bi-channel networks benefit each other to achieve superior performance with limited samples.

## III. EXPERIMENTS

### A. Dataset

We evaluated the proposed TSBN on a public INbreast dataset for classification of breast cancers [23], which contains 115 patients and 410 images. We assigned the samples of BI-RADS 1 and 2 as the negative cases, while those of BI-RADS 4, 5, and 6 were assigned as the positive cases. BI-RADS 3 denoted "probably benign", which was excluded in this work as the setting in [24] because of lacking reliable pathological confirmation. Therefore, the number of benign samples was 287, while that of malignant samples was 100.

### B. Experimental Setting

Table I shows the details of different comparison algorithms, including the pretext and downstream task networks, pretext tasks, and transfer strategies. In this work, U-Net and RDN were adopted as the pretext task networks for our GSIM task, respectively. The algorithm with U-Net was named TSBN-U, while that with RDN was named TSBN-R. To validate the effectiveness of the proposed TSBN, we compared the two algorithms with the following related classification algorithms:

- ResNet: The classical ResNet50 was selected as the baseline of classification network form comparison, which has been pre-trained on ImageNet for fine-tuning.
- SimCLR [25]: We adopted a typical fine-tuning based SSL algorithm SimCLR, which used instance discrimination as the pretext task with the ResNet50 as the backbone for the pretext and downstream tasks. This algorithm belonged to contrastive SSL.
- CR-F-ResNet [11]: We adopted a typical fine-tuning based SSL framework for comparison. This algorithm adopted the Context Restoration as a pretext task with the ResNet50 as the shared backbone network for both the pretext and downstream tasks. This algorithm belonged to generative SSL.
- GM-F-ResNet: This algorithm adopted the same fine-tuning based SSL framework and network architecture as CR-F-ResNet, but replaced the pretext task with the proposed Gray-scale Mapping task. This algorithm belonged to generative SSL.
- GM-M-ResNet: This algorithm adopted the conventional multi-task learning framework of hard parameter sharing for both the proposed Gray-scale Mapping and classification tasks [16], which were simultaneously conducted in a network with a shared ResNet50 backbone.

Table I
Details of different algorithms for SSL framework

| Algorithm | Pretext Task Network | Downstream Task Network | Pretext Task | Transfer Strategy |
|---|---|---|---|---|
| ResNet | / | / | / | Fine-tuning |
| SimCLR [25] | ResNet50 | ResNet50 | Instance Discrimination | Fine-tuning |
| CR-F-ResNet [11] | ResNet50 | ResNet50 | Context Restoration | Fine-tuning |
| GM-F-ResNet | ResNet50 | ResNet50 | GSIM | Fine-tuning |
| GM-M-ResNet | / | / | GSIM | Multi-task Learning |
| TSBN-U | U-Net | ResNet50 | GSIM | Collaborative Transfer |
| TSBN-R | RDN | ResNet50 | GSIM | Collaborative Transfer |

The 5-fold cross-validation strategy was applied to all algorithms in our experiment. The commonly used accuracy, sensitivity, specificity, Youden index (YI) were calculated as the evaluation indices. In addition, as an important metric for the problem of class imbalance, F1 was utilized for INbreast dataset in our experiment, because the number of positive cases is much less than that of negative cases. In addition, the receiver operating characteristic (ROC) and the corresponding area under its curve (AUC) were adopted for evaluation.

The mammograms were resized to 768×384 pixels. We adopted Adam optimization algorithm for each network with the learning rate 0.0001, weight decay 0.0005, and minibatch size 4.

*C. Results*

Table 2 shows the classification results of different algorithms. It can be found that both TSBN-U and TSBN-R outperform other compared algorithms, indicating the effectiveness of the proposed TSBN framework.

TSBN-U achieves the mean classification accuracy of 85.53±1.87%, sensitivity of 84.00±2.24%, specificity of 86.07±2.68%, YI of 70.07±2.84%, and F1 of 75.06%±2.52, which gets an improvement by 2.59%, 2.00%, 2.80%, 4.80% and 3.75% on accuracy, sensitivity, specificity, YI, and F1, respectively, over the ResNet. TSBN-U also gets an improvement by 1.07%, 2.00%, 0.75%, 2.75%, and 1.74% on accuracy, sensitivity, specificity, YI, and F1, respectively, over the typical contrastive SSL SimCLR, which indicates the effectiveness of the combination of the designed pretext task and the specific network.

In addition, TSBN-U improves at least 0.77% on accuracy, 2.00% on sensitivity and 2.38% on YI, and 1.38% on F1, compared with the conventional multi-task learning framework and the fine-tuning based generative SSL, including GM-M-ResNet, CR-F-ResNet, and GM-F-ResNet, indicating that more effective features representation are learned from the specific image restoration network. It also can be observed that GM-F-ResNet achieves the improvement of 1.26%, 1.69%, and 1.60% on accuracy, YI, and F1 over CR-F-ResNet, because the proposed GSIM task can provide more discriminative information than the classical image restoration task under the fine-tuning based SSL framework.

TSBN-R gets the same trend as TSBN-U, whose classification accuracy, sensitivity, specificity, and YI are 85.78±2.65%, 83.00±2.74%, 86.75±2.73%, 69.75±5.26%, and 75.18%, respectively. It improves at least 1.02% of accuracy, 1.69% of specificity, 2.06% of YI, and 1.50% of F1 compared with ResNet, SimCLR, CR-F-ResNet, GM-F-ResNet, and GM-M-ResNet, which indicates the effectiveness of our SSL framework and GSIM.

It can be concluded that since U-Net and RDN can generally perform image restoration tasks well, both networks are efficiently used in our SSL framework to conduct the proposed GSIM task. More effective knowledge is then transferred to the downstream classification network with improved performance.

Fig. 4 shows the ROC curves and corresponding AUC values of different classification algorithms. The proposed TSBN-U and TSBN-R achieve AUC values of 0.893 and 0.899, respectively. Both results are better than those of the fine-tuning based algorithms compared in this work, including contrastive SSL and generative SSL. These results again indicate the effectiveness of the proposed TSBN framework.

Table II
Classification results of different algorithms for mammography-based CAD on INbreast dataset (Unit: %)

| Methods | Accuracy | Sensitivity | Specificity | YI | F1 |
|---|---|---|---|---|---|
| ResNet | 82.94±2.17 | 82.00±4.47 | 83.27±1.99 | 65.27±5.48 | 71.31±3.52 |
| SimCLR | 84.46±2.57 | 82.00±2.74 | 85.32±3.83 | 67.32±3.50 | 73.32±3.10 |
| CR-F-ResNet | 83.48±2.68 | 82.00±2.74 | 84.00±4.51 | 66.00±2.09 | 72.08±2.72 |
| GM-F-ResNet | 84.74±3.30 | 82.00±2.74 | 85.69±3.66 | 67.69±6.03 | 73.68±4.77 |
| GM-M-ResNet | 84.76±2.08 | 81.00±2.24 | 86.07±2.41 | 67.07±3.87 | 73.36±3.04 |
| TSBN-U | 85.53±1.87 | **84.00±2.24** | 86.07±2.68 | **70.07±2.84** | 75.06±2.52 |
| TSBN-R | **85.78±2.65** | 83.00±2.74 | **86.75±2.73** | 69.75±5.26 | **75.18±3.92** |

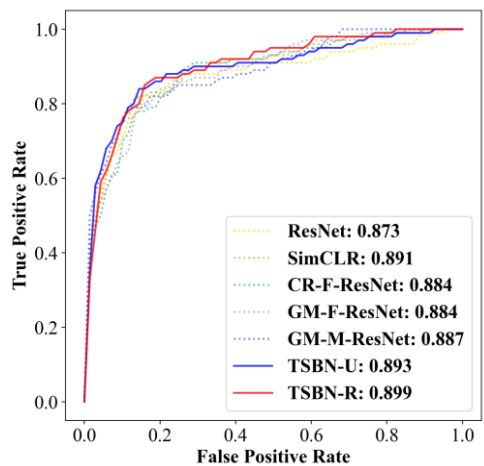

Fig. 4. ROC curves and corresponding AUC values of different algorithms for mammography-based CAD on INbreast dataset.

## IV. CONCLUSION

In this work, we propose a Task-driven Self-supervised Bi-channel Networks (TSBN) framework for the mammography-based CAD. In our SSL framework, a new GSIM task is designed as the pretext task, which can improve the feature representation with category-related information. The TSBN then innovatively integrates the specific image restoration network for GSIM and the downstream classification network into a unified SSL framework. The bi-channel networks corresponding to the pretext and downstream tasks are jointly trained, and the learned feature representations in both networks are collaboratively transferred to each other. This new SSL strategy not only improves the performance of downstream task, but also alleviates the issue of limited samples. The experimental results on the public INbreast dataset indicate that the proposed TSBN outperforms the conventional fine-tuning based SSL algorithms. In our future work, we will further explore ways to design different types of pretext tasks according to the label information and the characteristics of the medical data in the TSBN framework. In addition, we will also try to enhance the transfer performance between two different networks to further improve the classification accuracy of the downstream CAD model.

## ACKNOWLEDGMENT

This work is supported by the National Natural Science Foundation of China (81830058), the Science and Technology Commission of Shanghai Municipality (20ZR1419900), and the 111 Project (D20031).

## REFERENCES

[1] A.G. Waks and E.P. Winer, "Breast cancer treatment: a review," *Jama*, 2019, pp. 288-300.

[2] N.I.R. Yassin, S. Omran, E.M.F.El Houby, and H. Allam, "Machine learning techniques for breast cancer computer aided diagnosis using different image modalities: A systematic review," *Computer methods and programs in biomedicine*, 2018, pp. 25-45.

[3] L. Zou, S. Yu, T. Meng, Z. Zhang, X. Liang, and Y. Xie, "A technical review of convolutional neural network-based mammographic breast cancer diagnosis," *Computational and mathematical methods in medicine*, 2019.

[4] L. Tsochatzidis, L. Costaridou, and I. Pratikakis, "Deep learning for breast cancer diagnosis from mammograms—a comparative study," *Journal of Imaging*, vol. 5(3), pp. 37, 2019.

[5] M. Tariq, S. lqbal, H. Ayesha, I. Abbas, K.T. Ahmad, and M.F.K. Niazi, "Medical Image based Breast Cancer Diagnosis: State of the Art and Future Directions," *Expert Systems with Applications*, pp. 114095, 2020.

[6] L. Jing, and Y. Tian, "Self-supervised visual feature learning with deep neural networks: A survey," *IEEE Transactions on Pattern Analysis and Machine Intelligence*, 2020.

[7] X. Liu, F. Zhang, Z. Hou, Z. Wang, L. Mian, J. Zhang, J. Tang, "Self-supervised learning: Generative or contrastive," *arXiv:2006.08218 1.2*, 2020.

[8] M. Noroozi and P. Favaro. "Unsupervised learning of visual representations by solving jigsaw puzzles," *European Conference on Computer Vision*, Springer, 2016, pp. 69-84.

[9] D. Pathak, P. Krahenbuhl, J. Donahue, T. Darrell, A.A. Efros, "Context encoders: Feature learning by inpainting," *Proceedings of the IEEE conference on computer vision and pattern recognition*, 2016, pp. 2536-2544.

[10] A. Taleb, C. Lippert, T. Klein, and M. Nabi, "Multimodal self-supervised learning for medical image analysis," *International Conference on Information Processing in Medical Imaging*, Springer, 2021, pp. 661-673.

[11] L. Chen, P. Bentley, K. Mori, K. Misawa, M. Fujiwara, and D. Rueckert, "Self-supervised learning for medical image analysis using image context restoration," *Medical image analysis*, vol. 58, pp. 101539, 2019.

[12] Z. Zhou, V. Sodha, M.M.R Siddiquee, R. Feng, N. Tajbakhsh, M.B. Gotway, and J. Lian, "Models genesis: Generic autodidactic models for 3d medical image analysis," *International Conference on Medical Image Computing and Computer-Assisted Intervention*, Springer, 2019, pp. 384-393.

[13] H. Lee, S. .J Hwang, and J. Shin, "Self-supervised label augmentation via input transformations," *International Conference on Machine Learning*, 2020, pp. 5714-5724.

[14] A. Kolesnikov, X. Zhai, and L. Beyer, "Revisiting self-supervised visual representation learning," *Proceedings of the IEEE/CVF conference on computer vision and pattern recognition*, 2019, pp. 1920-1929.

[15] K. He, X. Zhang, S. Ren, and J. Sun, "Deep residual learning for image recognition," *Proceedings of the IEEE conference on computer vision and pattern recognition*, 2016, pp. 770-778.

[16] M. Crawshaw, "Multi-task learning with deep neural networks: A survey," arXiv:2009.09796, 2020.

[17] F. Zhuang, Z. Qi, K. Duan, et al, "A comprehensive survey on transfer learning," *Proceedings of the IEEE*, vol. 109(1). pp. 43-76, 2020.

[18] A. Khan, A. Sohail, U. Zahoora, and A.S. Qureshi, "A survey of the recent architectures of deep convolutional neural networks," *Artificial Intelligence Review*, 2020, pp. 5455-5516.

[19] O. Ronneberger, P. Fischer, and T. Brox, "U-net: Convolutional networks for biomedical image segmentation," *International Conference on Medical image computing and computer-assisted intervention*, Springer, 2015, pp. 234-241.

[20] Y. Zhang, Y. Tian, Y. Kong, B. Zhong, and Y. Fu, "Residual dense network for image super-resolution," *Proceedings of the IEEE conference on computer vision and pattern recognition*, 2018, pp. 2472-2481.

[21] X. Hu, M.A. Naiel, A. Wong, M. Lamm, and P. Fieguth, "RUNet: A Robust UNet Architecture for Image Super-Resolution," *Proceedings of the IEEE Conference on Computer Vision and Pattern Recognition Workshops*, 2019.

[22] N. Wu, J. Phang, J. Park, et al., "Deep neural networks improve radiologists' performance in breast cancer screening," *IEEE transactions on medical imaging*, 2019, pp. 1184-1194.

[23] I.C. Moreira, I. Amaral, I. Domingues, A. Cardoso, M.J. Cardoso, and J.S. Cardoso, "Inbreast: toward a full-field digital mammographic database," *Academic radiology*, 2012, pp. 236-248.

[24] L. Shen, L.R. Margolies, J.H. Rothstein, E. Fluder, R. McBride, and W. Sieh, "Deep learning to improve breast cancer detection on screening mammography," *Scientific reports*, 2019, pp. 1-12.

[25] T. Chen, S. Kornblith, M. Norouzi, and G. Hinton, "A simple framework for contrastive learning of visual representations," *Proceedings of International conference on machine learning*, 2020, pp. 1597-1607.